\documentclass[twocolumn,twocolappendix,tighten]{aastex631}
 \usepackage{mathptmx,courier}
  \usepackage[scaled=.92]{helvet}
  \usepackage{CJK}
  \DeclareMathAlphabet{\mathcal}{OMS}{cmsy}{m}{n}
\usepackage{natbib}
\usepackage{graphicx}
\usepackage[figuresright]{rotating}
\usepackage {amsmath,amssymb,eqnarray,tabularx}
\usepackage{enumerate}
\usepackage{enumitem}
\usepackage{mathtools}
\hypersetup{bookmarksnumbered=true,bookmarksopen=true}
\usepackage[all]{hypcap} 
\newcommand{\hi}{{\rm H}{\textsc i}}

\newcommand{\degree}{\ensuremath{\text{\textdegree}}}
\AtBeginDocument{
  \newcommand*{\cjkname}[3][gbsn]{#2 (\begin{CJK*}{UTF8}{#1}#3\end{CJK*})}
}

\def\kms{\rm{km~s^{-1} }}
\def\cmsq{\rm{ cm^{-2} } }
\def\NHI{N_{\rm HI}}
\def\MHI{M_{\rm HI}}

\def\Msunpcsq{\rm{M_{\odot}} ~{\rm pc}^{-2}}
\def\SHI{\Sigma_{\rm HI}}

\defcitealias{Wang23}{W23} 
\defcitealias{Wang24}{W24}

\begin{document}


\title{ FEASTS: Radial Distribution of HI surface densities down to $0.01\, \Msunpcsq$ of 35 Nearby Galaxies }
\correspondingauthor{Jing Wang}
\email{jwang\_astro@pku.edu.cn}

\author[0000-0002-6593-8820]{\cjkname{Jing Wang}{王菁}}
\affiliation{ Kavli Institute for Astronomy and Astrophysics, Peking University, Beijing 100871, China}

\author{\cjkname{Dong Yang}{杨冬}}
\affiliation{ Kavli Institute for Astronomy and Astrophysics, Peking University, Beijing 100871, China}

\author{\cjkname{Xuchen Lin}{林旭辰}}
\affiliation{ Kavli Institute for Astronomy and Astrophysics, Peking University, Beijing 100871, China}

\author{\cjkname{Qifeng Huang}{黄齐丰}}
\affiliation{ Kavli Institute for Astronomy and Astrophysics, Peking University, Beijing 100871, China}

\author{\cjkname{Zhijie Qu}{屈稚杰}}
\affiliation{Department of Astronomy and Astrophysics, The University of Chicago, 5640 S. Ellis Avenue, Chicago, IL 60637, USA}

\author{\cjkname[bsmi]{Hsiao-wen Chen}{陳曉雯}}
\affiliation{Department of Astronomy and Astrophysics, The University of Chicago, 5640 S. Ellis Avenue, Chicago, IL 60637, USA}

\author{\cjkname{Hong Guo}{郭宏}}
\affiliation{Shanghai Astronomical Observatory, Chinese Academy of Sciences, Shanghai 200030, People's Republic of China}

\author{\cjkname{Luis C. Ho}{何子山}}
\affiliation{ Kavli Institute for Astronomy and Astrophysics, Peking University, Beijing 100871, China}

\author{\cjkname{Peng Jiang}{姜鹏}}
\affiliation{National Astronomical Observatories, Chinese Academy of Sciences, 20A Datun Road, Chaoyang District, Beijing, China}

\author{\cjkname{Zezhong Liang}{梁泽众}}
\affiliation{ Kavli Institute for Astronomy and Astrophysics, Peking University, Beijing 100871, China}

\author{C{\'e}line P{\'e}roux}
\affiliation{European Southern Observatory, Karl-Schwarzschildstrasse 2, D-85748 Garching bei M{\"u}nchen, Germany}
\affiliation{Aix Marseille Universit{\'e}, CNRS, LAM (Laboratoire d'Astrophysique de Marseille) UMR 7326, F-13388 Marseille, France}

\author{Lister Staveley-Smith}
\affiliation{International Centre for Radio Astronomy Research, University of Western Australia, 35 Stirling Highway, Crawley, WA 6009, Australia}
\affiliation{ARC Centre of Excellence for All-Sky Astrophysics in 3 Dimensions (ASTRO 3D), Australia}

\author{Simon Weng}
\affiliation{Sydney Institute for Astronomy, School of Physics A28, University of Sydney, NSW 2006, Australia}
\affiliation{ARC Centre of Excellence for All-Sky Astrophysics in 3 Dimensions (ASTRO 3D), Australia}
\affiliation{ATNF, CSIRO Space and Astronomy,  PO Box 76, Epping, NSW 1710, Australia}

\begin{abstract}
We present the $\hi$ surface density ($\SHI$) radial distributions based on total-power $\hi$ images obtained by FAST in the FEASTS program, for 35 galaxies with inclinations lower than 72$\degree$. 
We derive the $\hi$ radius $R_{001}$, which is the radius for the 0.01 $\,\Msunpcsq$ ($\sim10^{18.1}\,\cmsq$) iso-density level, 100 times deeper than the 1 $\,\Msunpcsq$ level previously commonly used to measure $R_1$.
The profile shapes show a large diversity at a given radius in units of kpc, group virial radius, and $R_1$, but align more tightly with radius normalized by $R_{001}$.
The universal $\hi$ profile has a scatter of $\sim0.2$ dex, and a scale-length of $\sim0.11R_{001}$ in the outer region.
We derive a new $R_{001}$-$\MHI$ relation, which has a scatter of 0.02 dex, and similar slope of $\sim$0.5 as the previously known $R_1$-$\MHI$ relation. 
Excluding strongly tidal-interacting galaxies, the ratio $R_{001}/R_1$ (anti-)correlate strongly and significantly with the $\hi$-to-stellar mass ratio and sSFR, but not with the stellar mass, $\MHI$, dark matter mass, or SFR. 
The strongly tidal-interacting galaxies tend to show deviations from these trends, and have the most flattened profiles. 
These results imply that in absence of major tidal interactions, physical processes must cooperate so that $\SHI$ distributes in a self-similar way in the outer region down to the 0.01$\,\Msunpcsq$ level.
Moreover, they may drive gas flows in such a way, that $\hi$-richer galaxies have $\hi$ disks not only extend further, but also transport $\hi$ inward more efficiently from $R_{001}$ to $R_1$. 

\end{abstract}


\section{Introduction} 
\label{sec:introduction}
Correctly characterizing the inflow and outflow of gas has become a major bottleneck in developing galaxy evolution theory \citep{Crain23}.
The neutral atomic hydrogen ($\hi$) is a significant gas component fueling star formation, and promising tracer of the cool gas in and around galaxies.
Observing and combining it with multi-wavelength data is expected to provide key clues on gas flow processes. 

$\hi$ images are conventionally obtained with interferometric observation of the 21-cm emission lines, typically reaching a depth of $\sim10^{20}\,\cmsq$.
Based on them, it was revealed that $\hi$ disks show the following self-similar structures \citep{Wang16}. 
The relation between the $\hi$ mass and the semi-major axes of $\hi$ surface density ($\SHI$) contour at the $1\,\Msunpcsq$ ($\sim1.25\times10^{20}\,\cmsq$) level ($R_1$), is remarkably tight with a scatter of 0.06 dex.
The $\SHI$ radial profiles in the outer region show homogeneous shapes when the radius is normalized by $R_1$. 
These features are useful constraints to baryonic models in cosmological simulations \citep[e.g.,][]{Gensior24}.
One question is whether the self-similar behavior extends to lower surface densities.

Deeper than typical interferometric observations, detections of gas with $\NHI\sim10^{18}\,\cmsq$ is of significant importance to understanding baryonic flow and galaxy evolution.
It is the possible interface of disk confronting the circum-galactic medium (CGM; \citealt{FaucherGiguere23}), and frequently reveals optically dark interactions between galaxies (\citealt{Wang24}, hereafter W24), bearing important implications on the CGM structure and dynamics. 
Its coverage sets the mean free path of ionizing photons, regulating the photoionization states in the intergalactic medium (IGM). 
For decades, uncovering the faint $\hi$ regime has been relying on ultraviolet absorption observations, where large samples are required to overcome the incidental sampling limitation \citep{Borthakur15}.
Yet compiling a statistical sample in absorption for the $\sim10^{18}\,\cmsq$ regime, typically through Lyman Limit Systems (LLS), has also been challenging.
Latest developments consistently conclude on the complex environment, and therefore a statistically representative sample is required to characterize the connections between LLS properties and host galaxies \citep{Wotta19, Chen20, Peroux22, Weng23}.

Directly mapping $\hi$ with 21-cm emission line down to $10^{18}\,\cmsq$ for statistical samples is still difficult but has become possible in the nearby universe (\citealt{deBlok24}, \citetalias{Wang24}). 
It is exciting, not only for studies based on those data, but also because it provides reference to understand the behavior of LLSs at higher redshift.
It can be combined with LLSs to extract multi-resolution properties like the clumpiness or power spectrum.
It also bridges the disks to the even lower column density regime that further links the IGM, the Ly$\alpha$ absorption line detected regime of $\lesssim10^{15}\,\cmsq$. 

In this Letter, we use observations from FEASTS (FAST Extended Atlas of Selected Targets Survey, \citealt{Wang23}, hereafter W23) that directly maps $\hi$ down to a 3-$\sigma$ limit of $10^{17.7}\,\cmsq$.
Previous studies of FEASTS reveal that the $\hi$ of relatively isolated galaxies distributes in surprisingly flat disks at the data depth (Yang et al. in prep, \citetalias{Wang24}). 
This study thus focuses on the disk aspect of $\hi$ distribution: radial profiles of projection-corrected surface densities, and characteristic sizes based on them, for moderately inclined galaxies. 
Directly depicting $\hi$ disks at this new surface density limit, not only constrains disk formation, but also provides benchmark for further characterizing the column density distribution or coverage fraction, with random inclinations and environments for central galaxies. 
We assume a Kroupa IMF \citep{Kroupa01}, and do not include helium in calculations of the $\hi$ mass.

\section{Data and Method}
FEASTS is a long-term effort to obtain $\hi$ images for external galaxies selected by $\hi$ fluxes ($>$50 Jy km$\,{\rm s}^{-1}$), with the Five-hundred-meter Aperture Spherical radio Telescope (FAST, \citealt{Jiang19}) in the observable sky of FAST.
The data reduction is conducted with a customized pipeline, the details of which can be found in \citetalias{Wang23}. 
The data cubes have spatial resolution of 3.24$'$, and velocity resolution of 1.61 $\kms$. 
The source finding is performed with the software SoFiA \citep{Serra15}, with parameters listed in \citetalias{Wang24}. 
The parent sample used here is 55 galaxies already observed in the FEASTS program, during the period 2021.09-2024.06. 
The RFI (radio frequency interference) contamination incidence is less than 6.6\% for 95\% of the observations.
Assuming a line width of 20 $\kms$, the median 3-$\sigma$ depth in column density is 10$^{17.7}\,\cmsq$, corresponding to a face-on surface density of 0.004 $\Msunpcsq$.

The majority of the galaxies are within a distance of 20 Mpc, and have an integral $\hi$ flux above 100 Jy $\kms$. 
We take isophotal axis ratios measured in the Spitzer IRAC 1 (3.6 $\textmu$m) band from the S$^4$G catalog (The Spitzer Survey of Stellar Structure in Galaxies, \citealt{Sheth10}).
In the following analysis, we select from the starting sample the 35 not-too-inclined galaxies, which have IRAC-1 axis ratios larger than 0.3, roughly corresponding to inclinations lower than 72$\degree$.

\begin{figure*} 
\centering
\includegraphics{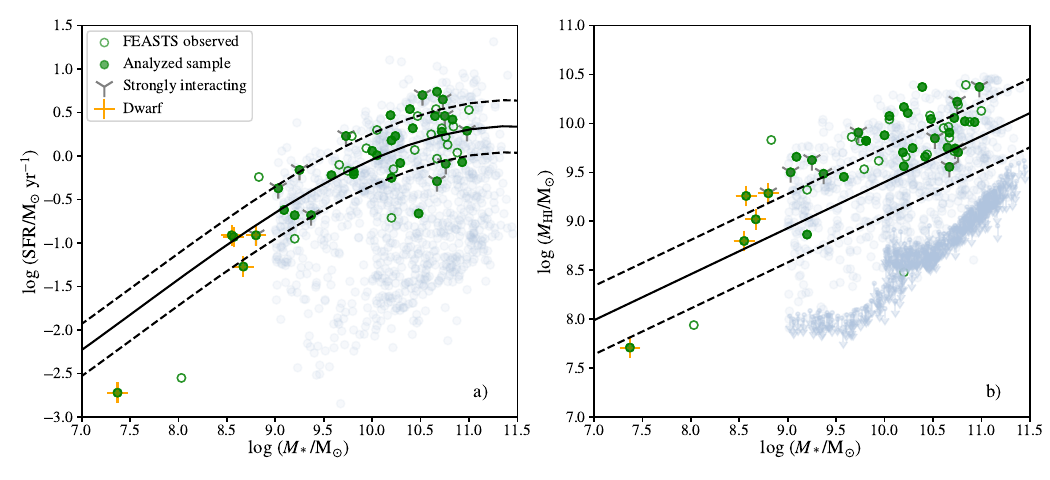}
\caption{Distribution of galaxies in the spaces of SFR and $\MHI$ versus $M_*$.
The green circles show the observed FEASTS sample, among which the filled ones are the subset analyzed here.
The grey Y-shape and orange X-shape symbols are for the strongly interacting and dwarf subsets respectively.
{\bf Panel a:} the star forming main sequence and related scatter from \citep{Saintonge16} are plotted as solid and dashed lines in the left panel.
{\bf Panel b:} the xGASS \citep{Catinella18} sample is displayed in the background as grey dots; its mean and scatter of $\MHI$-$M_*$ relation are plotted as solid and dashed lines.
}
\label{fig:ms}
\end{figure*}

The association of galaxies to galaxy groups, and group masses and virial radius ($r_{\rm vir}$) are taken from the group catalog of \citet{Kourkchi17}. 
Other properties including stellar masses ($M_*$), star formation rates (SFR), and distances are taken from the catalog of $z=0$ Multiwavelength Galaxy Synthesis (z0MGS, \citealt{Leroy19}). 
Figure~\ref{fig:ms} shows the distributions of analysis sample (and FEASTS-observed) galaxies in the 2D spaces of $\hi$ mass ($\MHI$) versus $M_*$, and SFR versus $M_*$.
There are 5 dwarf galaxies in the analysis sample having $M_*<10^9\, {\rm M}_{\odot}$, and others are MW-like. 
Compared to the general galaxy population at $z=0$, the sample is biased toward $\hi$-rich and star-forming galaxies, a compromise for observational convenience at this stage.

In a deep $\hi$ view, many galaxies experience some level of tidal interactions. 
For galaxies at a relatively early stage of merger, we perform 3D de-blending using a pipeline developed in \citet{Huang24} to separate the $\hi$ fluxes into individual galaxies (example atlas in Appendix A). 
This process is done on a best-effort basis, and biased against the galaxies with very close and small companions, or in late stage of merger.
Additionally, we specifically tag 11 most intensively interacting galaxies in the following analysis: IC 1727, NGC 672, NGC 3169, NGC 3368, NGC 4254, NGC 4449, NGC 4490, NGC 4532, NGC 4725, NGC 5033, NGC 5194. 
Readers can easily find numerous publications on each of these systems.
The remaining are less interacting galaxies.

We derive the $\hi$ surface density ($\SHI$) radial profiles for each galaxy using elliptical annulus around galaxy center with geometry set by the $\hi$ contour shape at $10^{20}\,\cmsq$. 
The axis ratios of the less interacting subset have 10, 50, and 90 percentiles of 0.51, 0.80, and 0.92.
We do not use the optical geometry, because $\hi$ disks typically warp outside the optical disks \citep{GarciaRuiz02, Kamphuis15, Wang17}, reflecting recent interactions with the external environment including the CGM.
Meanwhile, kinematical modeling for the outlying $\hi$ of the whole sample is being conducted and will be presented in a future paper.

The $\SHI$ values are deprojected by multiplying with the axis ratio ($\cos i$) of annulus, with a thin disk assumption.
Such a thin disk assumption may be less appropriate for dwarf irregular galaxies that are less confined by the disk gravity and more susceptible to perturbations \citep{Roychowdhury10}. We leave such a possible caveat to future quantification with high-resolution observations (e.g., with MeerKAT).
We derive characteristic radius at the $\SHI$ levels of 1 and 0.01$\,\Msunpcsq$, and refer to them as $R_1$ and $R_{001}$ \footnote{For three galaxies (all in the strongly interacting subset) that the radial profiles don't reach 0.01$\,\Msunpcsq$, localized linear extrapolation is conducted in the logarithm space to derive $R_{001}$.}.
The procedure is the same as the one used in \citep{Wang16}. 

We justify in Appendix~\ref{appendix:feasts} that these radius are robustly measured at the FAST resolution.
The typical deviations from true values due to beam smoothing are smaller than 0.1 dex for the range of $\hi$ disks angular sizes in this sample. 
There, we also describe procedures to correct for this small-extent bias, and estimate  uncertainties for $R_1$ and $R_{001}$ after taking into account both photometric error and beam smoothing effects.
The measurements of $R_1$ and $R_{001}$ together with errors are presented in Table~\ref{tab:size} in Appendix~\ref{appendix:size_table}. 

\section{Results}
\subsection{The $\SHI$ Radial Profiles}
\label{sec:profile}
We show all the $\SHI$ radial profiles in Figure~\ref{fig:profile}, with radius normalized to different units. 

In unit of kpc (Figure~\ref{fig:profile}-a), most MW-like galaxies have profiles extending beyond 50 kpc down to the data depth of $\log\,(\SHI/\Msunpcsq)=-2.4$.
The most extended of them go near 100 kpc.  
Possibly due to selection bias, these profiles are systematically higher than the median distribution predicted by simulations in \citet{Peroux20}.
The dwarf galaxies on average have the least extended profiles here, but do not differ significantly from the MW-like galaxies in other panels. 

Normalizing by the group virial radius $r_{\rm vir}$ reflects connection to the dark matter halo mass and CGM properties.
When normalized by $r_{\rm vir}$ (Figure~\ref{fig:profile}-b), the scatter of profiles is the largest among the first four panels. 
Galaxies have $r/r_{\rm vir}$ ranging between 0.05 to 0.5 when reaching the data depth, and the $\SHI$ at $r/r_{\rm vir}=0.2$ span $\gtrsim$3 index. 
These conclusions do not significant change, if we only select central galaxies ($20/24$ and $4/11$ for the less and strongly interacting subsets, respectively).

The $R_1$ and $R_{001}$ are characteristic radius for $\hi$ disks. 
When normalized by $R_1$ (Figure~\ref{fig:profile}-c), all $\SHI$ profiles align around the median profile derived with interferometric data \citep{Wang16} (W16 hereafter) in the inner region (within $R_1$).
Except for the strongly interacting galaxies, they are also close to the W16 median profile in the outer region, with offset $\lesssim$1.5 dex.
On the other hand, almost all profiles exceed the extrapolated W16 median profile in the outer region, reflecting the need to directly observe deep.
The strongly interacting galaxies show much more flattened shapes than less interacting galaxies in this normalization scheme.

Finally, when normalized by $R_{001}$ (Figure~\ref{fig:profile}-d), the profiles align in the tightest way throughout the radius. 
For less interacting galaxies, the scatters of $\SHI$ are $\lesssim$1 dex (or $\sigma\lesssim0.2$ dex) at most radius. 
Even the profiles of the strongly interacting galaxies do not deviate far in this normalization scheme.
The $\SHI$ distribution thus looks self-organized: it is most strongly determined by its own largest extension, rather than by plain unit (e.g., kpc) scales or the dark matter halo size. 
Such a self-organization is remarkable, when one considers that galaxies are ecosystems, CGM are multiphase, and $\hi$ gas is susceptible to perturbations. 
We are not inferring that effects from other baryonic components, dark matter or radiation field do not influence the $\hi$ distribution.
Instead, their net effect produces the self-similar $\SHI$ distribution all the way down to the 0.01 $\Msunpcsq$ level, which is a strong constraint to galaxy formation models.

\subsubsection{Analytical Approximations of the Median $\SHI$ Profile}
For future comparison convenience, the median profile of less interacting galaxies is provided in Table~\ref{tab:medprof}, and plotted again in Figure~\ref{fig:approx_profile}.
The median profile has an exponential shape on the outer region with scale-length $r_s=0.11R_{001}$ (purple line in Figure~\ref{fig:approx_profile}, derived with the average $R_{001}/R_1$ ratio in Section~\ref{sec:size}).

We use the analytical equational form from \citet{Wang14} to fit the shape of the median profile, and obtain the following (blue curve in Figure~\ref{fig:approx_profile}):
\begin{equation}
\label{eq:expapprox}
y=\log \frac{I\, e^{-x/r_{s1} }}{1+ (I/J-1)e^{-x/r_{s2} } }
\end{equation}
where $y=\log\, (\SHI/\Msunpcsq)$, $x=r/R_{001}$, $I=62.95^{+14.78}_{-11.13}$, $J=5.96^{+0.34}_{-0.32}$, $r_{s1}=0.114^{+0.004}_{-0.004}$, and $r_{s2}=0.106^{+0.007}_{-0.006}$. 
This equation form is based on the previous observational findings that both total neutral gas surface densities ($I$) and the $\hi$-to-H$_2$ conversion efficiencies (${\rm H}_2/\hi=(I-J)/J$) decrease with radius roughly following exponential functions \citep{Leroy08, Bigiel08, Bigiel10}.
It has been found to well fit the individual $\SHI$ profiles of MW-type galaxies at $\SHI\gtrsim 0.5 \Msunpcsq$ levels \citep{Wang14}.
Here it also well describes the median $\SHI$ profile down to the $\lesssim0.01\,\Msunpcsq$ level.

The overall shape can also be roughly described by the following cored power law equation with two less parameters (green curve in Figure~\ref{fig:approx_profile}):
\begin{equation}
  \label{eq:polyapprox}
y=\log \frac{ (1+(x/r_c)^2)^{-\beta}}{100 (1+(1/r_c)^2)^{-\beta} }
\end{equation}
where $r_c=0.94^{+0.09}_{-0.07}$, and $\beta=8.43^{+1.15}_{-0.87}$.


\subsubsection{The Curves of Growth of HI mass}
\begin{figure*} 
  \centering
  \includegraphics{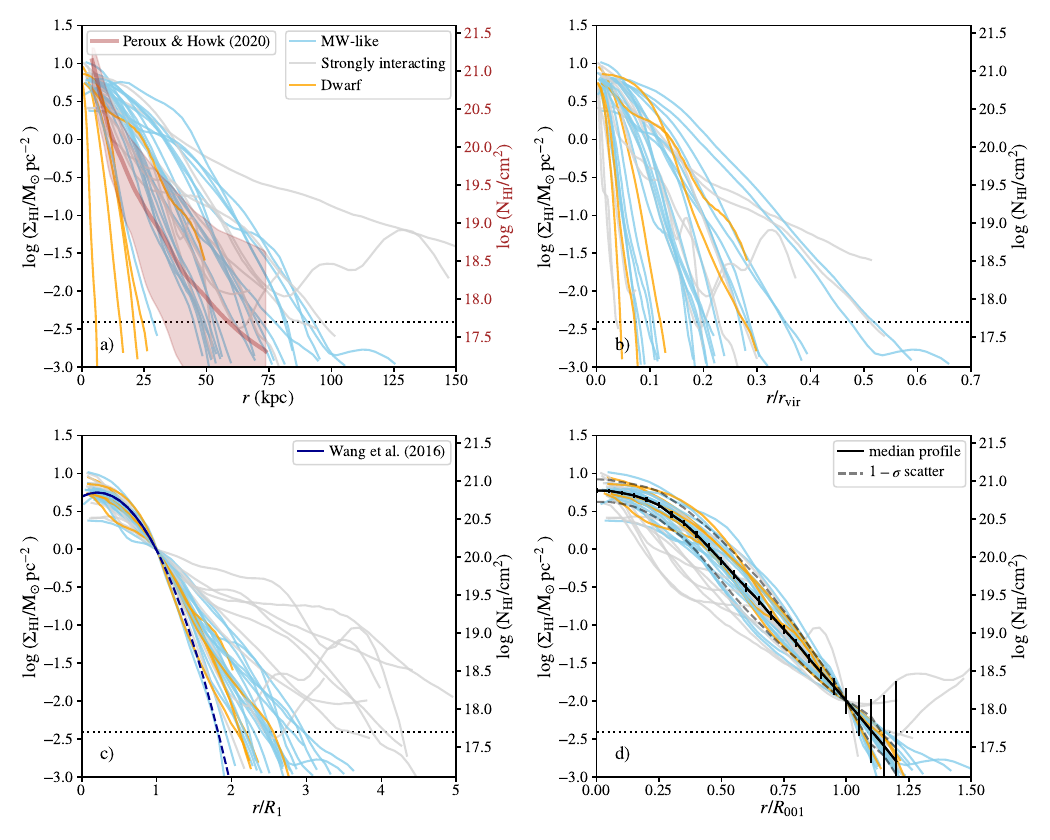}
  \includegraphics{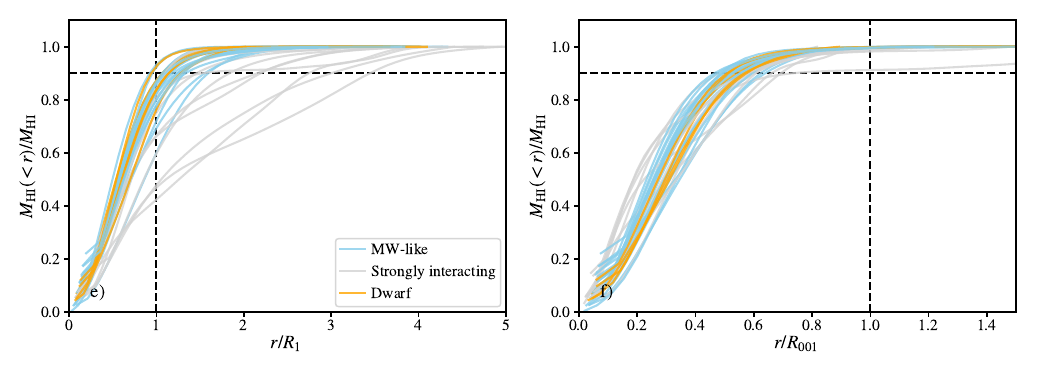}
  \caption{Radial profiles of $\hi$ distribution. The blue, orange and grey colors are for the MW-like, dwarf and strongly interacting galaxies, respectively. {\bf Panels a-d:} $\SHI$ profiles with radius in units of kpc, $r_{\rm vir}$, $R_1$, and $R_{001}$. The dashed horizontal lines mark the data depth of $\log\,(\SHI/\Msunpcsq)=-2.4$. The brown shaded region in panel a shows simulation predicted typical $\hi$ column density profile of MW-type galaxies (from Figure 1 of \citealt{Peroux20}). The black solid curve in panel c shows the interferometry data-based $\SHI$ median profile from \citet{Wang16}, with the dotted curve being its extrapolation. The black solid curve in panel d shows the median $\SHI$ profile for less interacting galaxies here, and the dashed curves the 1-$\sigma$ scatter. {\bf Panels e and f:} normalized $\hi$ mass curve-of-growth with radius in units of $R_1$, and $R_{001}$. The dashed horizontal and vertical lines mark the positions of 0.9 and 1, respectively. }
  \label{fig:profile}
\end{figure*}

\begin{figure} 
  \centering
  \includegraphics{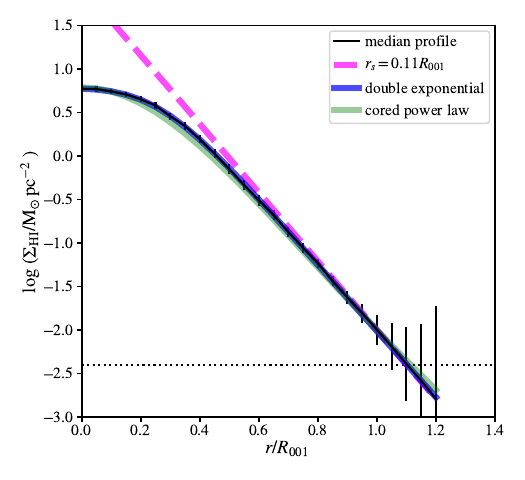}
  \caption{Mathematical approximations for the $R_{001}$-normalized $\SHI$ median profile. The median profile and dotted line are the same as in Figure~\ref{fig:profile}. The green curve shows the polynomical approximation of Equation~\ref{eq:polyapprox}, and the magenta dashed line the exponential approximation for the outer region (Section~\ref{sec:size}).}
  \label{fig:approx_profile}
\end{figure}

In the bottom row (panels e and f) of Figure~\ref{fig:profile}, we also show how the $\hi$ masses cumulatively increase with radius in views normalized by $R_1$ and $R_{001}$ respectively. 
It is clear that, the $\MHI$ are few ($<$20\%) beyond $R_1$, and minimal beyond $R_{001}$. 
Thus, previous studies based on interferometry observations have already captured the bulk $\hi$ masses (assuming insignificant short-spacing problems).
The focus of this study is therefore not on where most of the $\hi$ lies, but on the diagnostic power in the distribution of this low-surface-density and wide-spreading $\hi$ on $\hi$-CGM interaction problems. 
This minimal amount of $\hi$ between $R_1$ and $R_{001}$ lies in the disk-CGM interface where inflows and outflows actively happen and determine the galaxy total $\MHI$. 
We will further elaborate this point in Section~\ref{sec:size}.

\subsection{A New $\hi$ Size-mass Relation at the 0.01$\,\Msunpcsq$ level}
\label{sec:size}
\begin{figure} 
  \centering
  \includegraphics{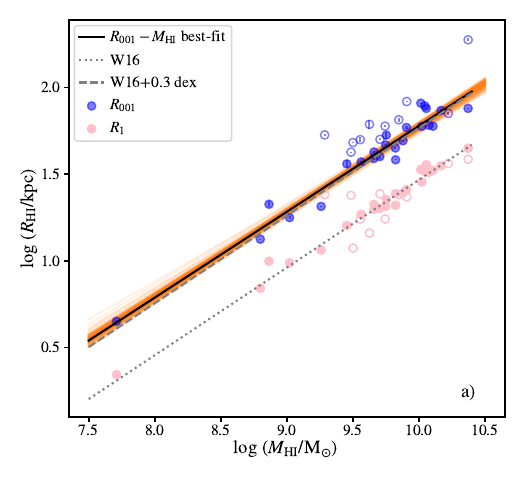}
  \caption{ Size-mass relation of $\hi$. The blue and pink symbols are for $R_{001}$ and $R_1$ measurements respectively. The open circles are for strongly interacting systems. The solid line plots the $R_1$-$\MHI$ relation from \citet{Wang16}, and the dashed line is the solid line moving upward by 0.3 dex. The black solid line shows the best-fit $R_{001}$-$\MHI$ relation, for which the {\it emcee}-based uncertainties are displayed as transparent orange lines. 
 }
  \label{fig:radius}
\end{figure}

The previously reported $R_1$-$\MHI$ relation was based on interferometry data \citep{Swaters02, Wang16}, suffering from the possible problem of short-spacing or missing-flux.
In Figure~\ref{fig:radius}, the $R_1$ and $\MHI$ measured here, which do not have missing fluxes problem, follow well the previous $R_1$-$\MHI$ relation. 
It suggests that either the levels of underestimate in previous studies due to missing-flux are low for most interferometric observations, or just move galaxies along the relation.  

Figure~\ref{fig:radius} shows that $R_{001}$ and $\MHI$ also have a strong correlation with Pearson R value of 0.92. 
It is consistent with the homogeneous $R_{001}$-normalized profiles in Figure~\ref{fig:profile}.
In contrast, the correlations of $R_{001}$ with $M_*$ or $M_{\rm halo}$ are weaker, with R values of 0.80 and 0.56, respectively. 
The strongly interacting galaxies have 1.4 times systematically higher $R_{001}$ at a given $\MHI$ than the less interacting ones, in contrast to their $R_1$ being scattered around the $R_1$-$\MHI$ relation.
If we shift the $R_1$-$\MHI$ relation upward by 0.31 (0.34) dex, most $R_{001}$ versus $\MHI$ data points excluding (including) strongly interacting galaxies lie on it. 
So at a given $\MHI$, $R_{001}$ is around 2 (2.2) times $R_1$. 
For less interacting systems, the ratio $R_{001}/R_1=2$ indicates an outer profile scale-length $r_s=0.22R_{1}=0.11R_{001}$ (Figure~\ref{fig:profile}-d).

We conduct a linear regression for the less interacting subset, and derive the following relation:
\begin{equation}
\log R_{001}=0.49(\pm0.02)\log \MHI -3.11 (\pm0.21)
\end{equation}
The {\it emcee} method \citep{ForemanMackey13} is used to derive the slope and intercept uncertainties, as well as a fractional intrinsic scatter $<0.001$ of the relation.
The standard deviation of data points around the relation is 0.02 dex, slightly higher than the 0.01 dex around the $R_1$-$\MHI$ relation with the same sample.
Such a tight relation can be used to estimate $R_{001}$ based on the $\MHI$, which will be useful for studies where deep $\hi$ images are not always available. 

Like the $R_1$-$\MHI$ relation, the new $R_{001}$-$\MHI$ relation has a slope close to 0.5.
It suggests an almost constant average surface density of $\hi$ within $R_{001}$, determined by the intercept $b$ of the $R_{001}$-$\MHI$ relation with a value $\Sigma_{\rm HI,avg}=10^{-2b}/\pi=0.35\, \Msunpcsq$. 
It also suggests that distance uncertainties only shift galaxies along the relation and have a minimal effect on the scatter.

For the less interacting galaxies, while the area between $R_1$ and $R_{001}$ is 3 times larger than that within $R_1$, most (80\%) $\MHI$ are enclosed within $R_1$ (Figure~\ref{fig:profile}-e), and the majority (90\%) of them are within 0.5$R_{001}$ (Figure~\ref{fig:profile}-b).
It is thus the structure instead of $\hi$ mass budget that makes the $R_{001}-\MHI$ relation and $R_{001}$-normalized profiles interesting.  
This small amount of $\hi$ mass beyond $R_1$ (or between 0.5 and 1 $R_{001}$) could have distributed and ended anywhere, yet it seems to almost always reach 0.01$\, \Msunpcsq$ at $\sim2R_1$ (with scatter$<$0.1 dex).
We test the relation of $\hi$ mass between $R_1$ and $R_{001}$ with the area $\pi(R_{001}^2-R_{1}^2)$ and volume $\frac{4}{3}\pi(R_{001}^3-R_{1}^3)$.
The former has smaller scatter (0.12 dex) than the latter (0.18 dex), supporting a disky instead of spherical $\hi$ distribution.

\subsection{The Dependence of $R_{001}/R_1$}
\label{sec:ratio}
As both $R_{001}$ and $R_1$ have tight relations with $\MHI$, it is useful to ask whether any new clues are revealed by $R_{001}$ on building the $\MHI$.
Because the $\SHI$ profiles are well aligned locally when normalized by either $R_1$ or $R_{001}$, we ask, whether the profile slope between these two radius carry information on regulation from properties other than $\MHI$. 
Both questions can be studied by looking into how the size ratios, $R_{001}/R_1$, relate to other galaxy properties. 

\begin{figure*} 
\centering
\includegraphics[width=6.5cm]{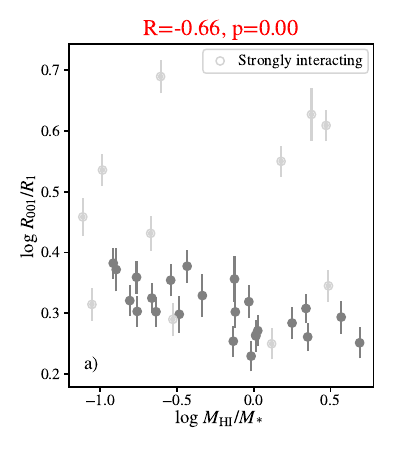}
\includegraphics[width=6.5cm]{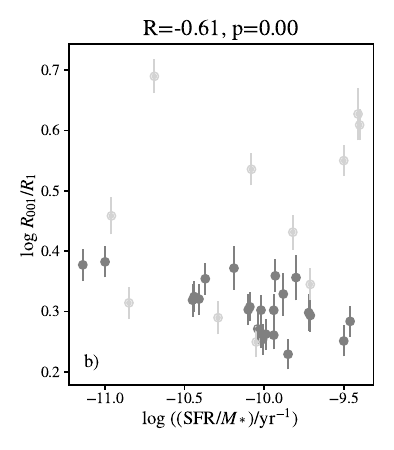}
\includegraphics[width=6.5cm]{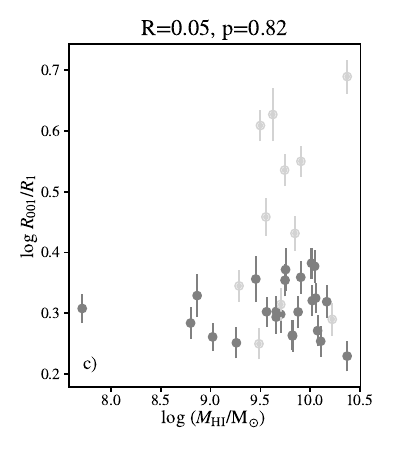}
\includegraphics[width=6.5cm]{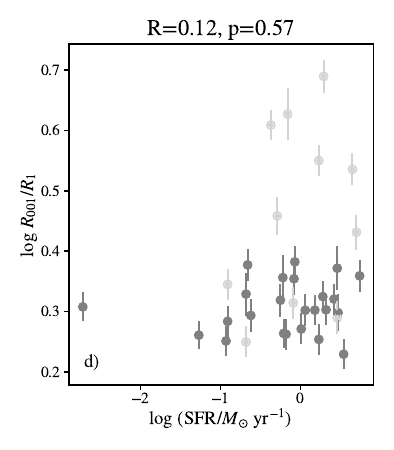}
\includegraphics[width=6.5cm]{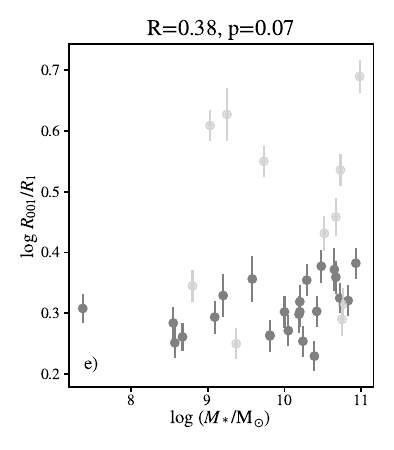}
\includegraphics[width=6.5cm]{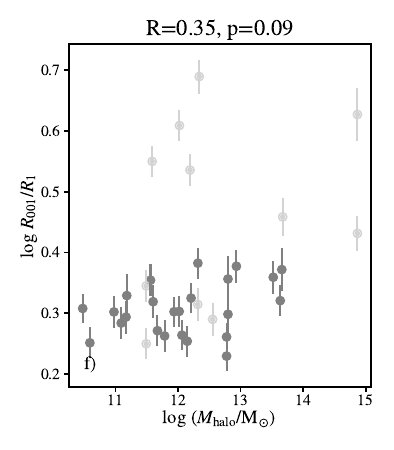}
\caption{The dependence of $\hi$ size ratio $R_{001}/R_{1}$ on galaxy properties. The galaxy properties examined include the $\hi$ richness (a), the specific SFR (sSFR, b), $\MHI$ (c), $M_*$ (d), SFR (e), and group halo mass ($M_{\rm halo}$, f). The Pearson R and p values of correlation for less interacting galaxies are denoted on top of each panel. The strongest correlation with $\MHI/M_*$ is highlighted in red.
The strongly interacting galaxies are marked with open circles.  }
\label{fig:ratio}
\end{figure*}

An investigation of these relations is presented in Figure~\ref{fig:ratio}.
The Pearson $R$ and $p$ values are used to quantify the correlation strength and significance of each relation, as labeled in each panel. 
We confirm that switching to the Spearman correlation coefficients does not significantly change the following results.
As a further robustness check, for correlations identified to be strong and significant, we also derive the partial correlation coefficients, removing possible mutual dependence of $R_{001}/R_1$ and parameters on galaxy distances.

Figure~\ref{fig:ratio} (panels a and b) shows that the $R_{001}/R_1$ ratios depend most strongly and significantly on the $\hi$ mass fraction ($\MHI/M_*$) and secondly on specific SFR (SFR$/M_*$). 
As the slopes of the two size-mass relations are close, $R_{001}/R_1$ have a rather small dynamic range ($\lesssim0.2$\,dex), yet they show clear trends with these two parameters. 
The partial correlation coefficients controlling for distances suggest that the $R_{001}/R_1$-$\MHI/M_*$ correlation is highly robust (Pearson $R=-0.55$, $p=0.01$), while the $R_{001}/R_1$-sSFR correlation becomes considerably weaker and insignificant (Pearson $R=-0.23$, $p=0.30$). 
We reach a similar conclusion when testing the partial correlation between $R_{001}/R_1$ and $\MHI/M_*$ (sSFR) with the effect of sSFR ($\MHI/M_*$) being controlled for. 
Thus, the $R_{001}/R_1$-$\MHI/M_*$ relation seems to be the most important and possibly most intrinsic one among the relations studied here.

Figure~\ref{fig:ratio} (panels c to f) also shows that $R_{001}/R_1$ do not exhibit strong or significant dependence (reflected in either low R or high p values) on the $\hi$ mass, stellar mass, dark matter halo mass, or SFR, although these parameters are considered important in determining galaxy evolution. 

We note that removing the five dwarf galaxies does not significantly change the results presented above.
Therefore, the steepness of $\SHI$ profiles in outer regions indeed point to effects other than self-organization of $\hi$ disk. 
Such that, gas-richer galaxies tend to have steeper $\SHI$ profiles with respect to $R_1$ in the outer regions. 

Finally, while the highest values of $R_{001}/R_1$ value are contributed by the strongly interacting galaxies, only 4 out of 11 strongly interacting galaxies stand out as clear outliers in the $R_{001}/R_1$-$\MHI/M_*$ relation. 
Together with the result of strongly interacting galaxies fluctuating around the $R_1-\MHI$ relation, it implies some global and regular response of $\hi$ outer disks to external perturbations, possibly also linking to self-regulation via interacting with the CGM.

\section{Summary and Discussion}
\label{sec:summary}
It was found that galaxies with higher SFR at a given $M_*$ have a higher $\MHI$ \citep{Saintonge22}, higher fraction of $\hi$ mass beyond the optical disk over the total $\hi$ mass \citep{Wang20}, and higher incidence and strength of cool gas absorbers in the CGM \citep{Dutta24}.
These relations are direct constraints to the quasi-equilibrium states of baryonic flow through galaxies, in which the relative rates of gas accretion, gas inflow, star formation, and gas outflow determine the baryonic mass reservoir in each step, the so-called gas-regulator or bathtub model \citep{Bouche10, Dave12, Lilly13}. 
One possible interpretation is that for star-forming galaxies gas accretion is highly efficient on large scales, whereas inward flows within the galaxy disk are comparatively less effective.
In other words, while the specific rates of both gas accretion (onto the $\hi$ disk) and inflow (radially through the $\hi$ disk and into the optical radius) increase with sSFR, the former possibly increases more quickly than the latter. 
So that, a large reservoir of $\hi$ can be built up beyond the optical disks of actively star-forming galaxies.

Our results provide new details to the aforementioned picture, specifically on the gas flow between $R_1$ and $R_{001}$. 
The tight relation between $\MHI$ and $R_1$ has been taken as an important characterization of the $\hi$ disk structure, and strong constraint on the built-up and shrinking of $\hi$ disks as a reservoir of star-forming material in galaxy formation models (e.g., \citealt{Gensior24}).
The relation between $\MHI$ and $R_{001}$ confirms that this reservoir in the form of $\hi$ disks structurally extends to the $\SHI$ level of $0.01\,\Msunpcsq$. 
Importantly, these $\hi$ radius being most tightly related with $\MHI$ infers that the balance of inflow and outflow organizes the $\hi$ distribution in a self-similar way out to to $R_{001}$. 
It is consistent with recent finding that Ly$\alpha$ absorber strengths within the virial radius correlate more strongly with $\MHI$ than with $M_*$, SFR, or $M_{\rm vir}$ \citep{Borthakur24}. 
But our results for the first time directly quantify $\SHI$ as a function of $r$, and uniquely indicate that at $\SHI>0.01\,\Msunpcsq$ this reservoir tends to take the format of a flat disk, instead of halo gas, because the disk structure is assumed and preferred here to obtain the tight relations (also see \citetalias{Wang24} and D. Yang in prep for direct images of face-on and edge-on disks).
The trend of $R_{001}/R_1$ decreasing with $\hi$-richness (sSFR) for less interacting galaxies implies that the gas flow from $R_{001}$ to $R_1$ is efficient, possibly because the newly condensed gas at large radius is not yet fully supported by rotational velocity \citep{Stern21}. 
Hence, the aforementioned bottleneck of gas inflowing to the optical disk possibly happens inward $R_1$.
The possibly relatively fast and slow inflows near $R_{001}$ and optical radius respectively naturally lead to cumulation of gas between these two radius, which may explain why higher fractions of $\hi$ mass beyond optical disks tend to be found around more star-forming galaxies \citep{Wang20}.

\begin{figure*} 
  \centering
  \includegraphics[width=\textwidth]{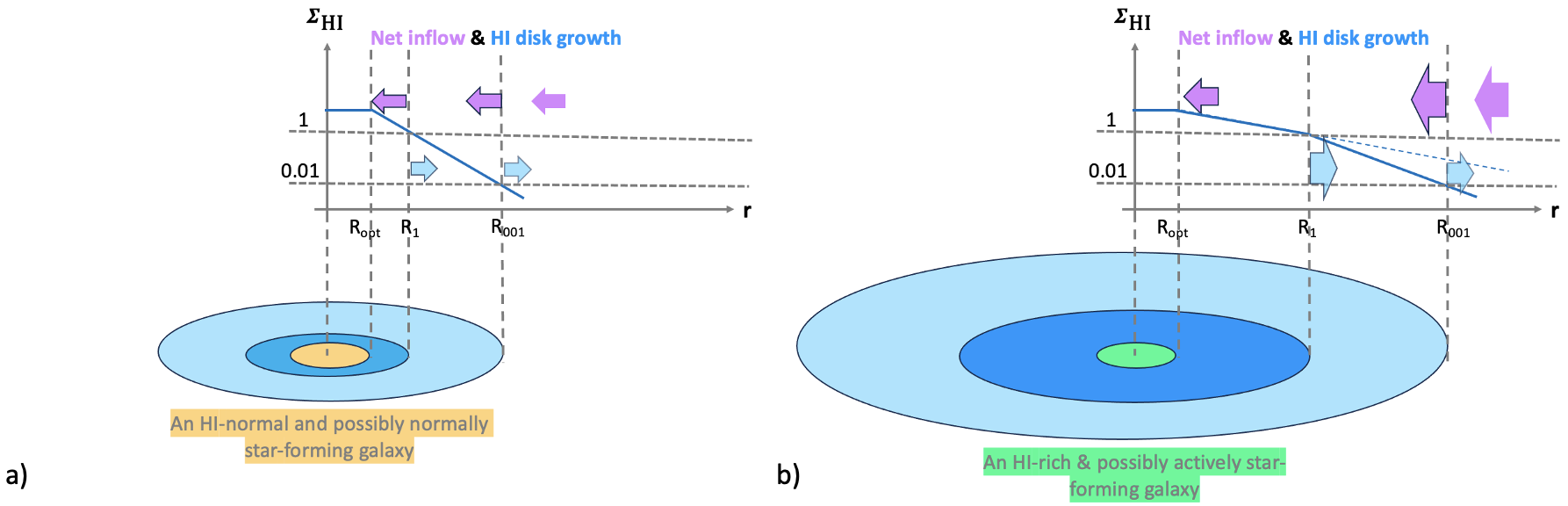}
  \caption{ A sketch on how net inflow of gas may build up the $\hi$ disks for galaxies with different $\hi$-richnesses. 
  Panels a and b are for an $\hi$-normal and an $\hi$-rich galaxy respectively.  
  The blue curves represent the $\SHI$ radial profiles of the corresponding galaxies, with the $\hi$ radius $R_{001}$ and $R_1$, and the optical radius $R_{\rm opt}$ marked on the x-axis. 
  The blue dashed curve in panel b represents a log-linear extension of the profile within $R_1$, so indicates the down-bending of the actual profile toward $R_{001}$. 
  The purple arrows represent the net $\hi$ inflow, and the blue arrows represent the $\hi$ disk inside-out growth as a result of net inflows.  
  The widths of purple and blue arrows indicate the relative (not to scale) strengths of corresponding processes at different radii and between the two types of galaxies.  
  See the third paragraph of Section~\ref{sec:summary} for more details.
 }
  \label{fig:scenario}
\end{figure*}

We use a sketch in Figure~\ref{fig:scenario} to help explain the scenario discussed above.
In the sketch,  the two example galaxies have the same optical disk size. 
While the $\hi$-rich galaxy has larger $R_1$ and $R_{001}$, it has smaller $R_{001}/R_1$ compared to the $\hi$-normal one. 
 The purple arrows represent the net $\hi$ inflow: the rightmost arrow indicates inflow from the CGM onto the extended $\hi$ disk (possibly from a different gas phase and different structure from the $\hi$ disk), the middle arrow shows the flow of gas from $R_{001}$ to $R_1$, and the leftmost arrow represents the flow from $R_1$ to within the galaxy optical disk. 
 Within each galaxy, the $\hi$ is accreted onto the $\hi$ disk near $R_{001}$ in a faster way than it enters the optical disk, so that $\hi$ piles up beyond $R_{\rm opt}$ and the $\hi$ disk is larger than the optical disk. 
 Compared to the $\hi$-normal galaxy, the $\hi$-rich galaxy has enhanced inflows at all radii, but the inflow into the $\hi$ disk at $R_{001}$ is enhanced more than that into the optical disk at $R_{\rm opt}$, so that its $\hi$ disk grows in size more significantly, resulting in higher $R_{001}/R_{\rm opt}$ and higher $R_1/R_{\rm opt}$ (i.e., $\hi$-richer galaxies have larger optical-to-$\hi$ size ratios and higher fraction of $\hi$ beyond the optical disks, \citealt{Wang14, Wang20}).
  Within the $\hi$-rich galaxy, the $\hi$ flows from $R_{001}$ to $R_1$ in a faster way than it from $R_1$ into the optical disk, so as a result, the $\hi$ disk grows faster near $R_1$ than it does near $R_{001}$, leading to a down-bending $\hi$ disk and smaller value of $R_{001}/R_1$ compared to the $\hi$-normal disk (i.e., $\hi$-richer galaxies have lower $R_{001}/R_1$ values).
  
  Because the localized $\SHI$ is the result of balance between cooling, heating, ionization, radial inflow, outflow, and stripping, an alternative interpretation for the trend of $R_{001}/R_1$ decreasing with $\MHI/M_*$ can be truncation in the outer disks due to ionization or dynamic effects. 
However, the $R_{001}/R_1$-$R_{001}/r_{\rm vir}$ correlation is considerably weaker (Pearson $R=-0.13$ and $p=0.55$) than the $R_{001}/R_1$-$\MHI/M_*$ correlation, which seems to disfavor this picture. 
Velocity field modeling and multiphase-gas observation in the outer disks will be essential to directly differentiate the possible scenarios.
Existing (e.g., MHONGOOSE) and future high-resolution deep $\hi$ observations may be efficient at directly detecting signatures of gas accretion near $R_{001}$. 

These observational details provide useful clues on CGM dynamics and gas accretion in galaxy formation models. 
Below we give an example of qualitatively comparing with the cooling flow model proposed with the FIRE-2 simulation \citep{Hopkins18}. 
This model suggests subsonic cooling flows conserving angular momentum to be a feasible way of gas accretion in MW-type halos \citep{Stern19}. 
One key prediction is that hot gas only condenses near the ``circular radius'' determined by the virial radius and average halo spin \citep{Stern21}, and thus forming a flat, well-aligned, rotating cool gas disk \citep{Hafen22}.  
The related simulations also predict that the accreted gas tends to pile around $R_1$, which effectively defines the disk edge \citep{Trapp24}.
The $R_{001}$-$\MHI$ relation qualitatively supports this model on the aspect of gas accretion directly building an $\hi$ disk, instead of precipitating into a halo of cool clouds.
But the relation and homogeneously $R_{001}$-normalized profiles also indicate that the newly accrete gas does not pile up around the same circular radius or $R_1$, but seems to smoothly ``move'' outward.
The moving out is in a way corresponding only to the existing total $\hi$ mass, but not to the virial radius or halo spin.
The anti-correlation between $R_{001}/R_1$ and gas richness implies that the edge of the disk may be beyond $R_{001}$, rather than around $R_1$. 
We expect the results to be also useful to constraining other models and simulations.

Finally, we comment briefly on the strongly interacting galaxies, which tend to have much more extended $\hi$ distribution than other galaxies.
Such distribution increases the interface of $\hi$ with CGM, and thus possibly localized CGM cooling rate through turbulent mixing \citep{Sparre22}. 
Previous observations did find statistical evidence for post-mergers to show enhanced $\hi$ richness \citep{Ellison18}, and localized cooling enhancement is consistent with $\hi$ distribution and kinematics in some individual interacting systems (\citetalias{Wang24}, X. Lin in prep). 
However, the universality and exact stage of that happening remains to be quantified. 

The results of our work can be re-examined using the up-coming high-resolution and deep observations of moderately-large nearby galaxies at MeerKAT, ngVLA, and SKA, in the future.

\section*{acknowledgments}
We thank the anonymous referee very much for constructive comments. 
We also gratefully thank Filippo Fraternali, W.J.G. de Blok, Luca Cortese for useful discussions.
JW thanks support of the research grants from Ministry of Science and Technology of the People's Republic of China (NO. 2022YFA1602902),  the National Natural Science Foundation of China (NO. 12073002), and the science research grants from the China Manned Space Project (NO. CMS-CSST-2021-B02).  
LCH was supported by the National Natural Science Foundation of China (11721303, 11991052, 12011540375, 12233001), the National Key R\&D Program of China (2022YFF0503401).
Parts of this research were supported by the Australian Research Council Centre of Excellence for All Sky Astrophysics in 3 Dimensions (ASTRO 3D), through project number CE170100013.
Parts of this research were supported by High-performance Computing Platform of Peking University.

This work made use of the data from FAST (Five-hundred-meter Aperture Spherical radio Telescope, \url{https://cstr.cn/31116.02.FAST}). FAST is a Chinese national mega-science facility, operated by National Astronomical Observatories, Chinese Academy of Sciences.

\facilities{FAST: 500 m, VLA }
\software{Astropy \citep{astropy:2022}, 
numpy \citep[v1.21.4]{vanderWalt11}, photutils \citep[v1.2.0]{Bradley19}, Python \citep[v3.9.13]{Perez07}, scipy \citep[1.8.0]{Virtanen20} }

\appendix   
\section{Deblending HI disks Undergoing Interactions}
We follow similar procedures outlined in \citet{Huang24} to perform 3D source de-blending and separate the $\hi$ flux in the interacting systems. 
It uses optical information as initial guess for positions of galaxies, and utilizes the \textit{watershed} algorithm in the 3D to segment a data cube into individual galaxies.

In brief, we perform a two-pass \textit{watershed} on the FEASTS data cubes. 
During the first pass, initial markers are set at voxels contained in the optical disk for each galaxy (i.e., the sky coordinates of the markers are the same in all channels).
The \textit{watershed} algorithm is then executed based on these markers to de-blend the data cubes. 
In the second pass, we identify the local maxima of $\hi$ flux density within the data cube region of each galaxy that has been allocated in the first pass, and utilize them to update the markers. 
The \textit{watershed} algorithm is run again to achieve the final segmentation of the data cubes.
In \citet{Huang24}, we use mock test to show that this procedure successfully segments HI data cubes of merging pairs that have projected distances larger than $0.5 (R_{1,1}+R_{1,2})$, or radial velocity differences larger than $0.5(W_1+W_2)$. 
The $R_{1,1}$ and $R_{1,2}$ are the $R_1$, and the $W_1$ and $W_2$ are the integral $\hi$ line width of the two galaxies in each pair.
The uncertainties in $\hi$ flux for the deblended primary galaxies are less than 10\%. 
And the procedure works significantly better than deblending galaxies on the 2D moment maps.

\begin{figure*}
    \centering
    \includegraphics[width=\linewidth]{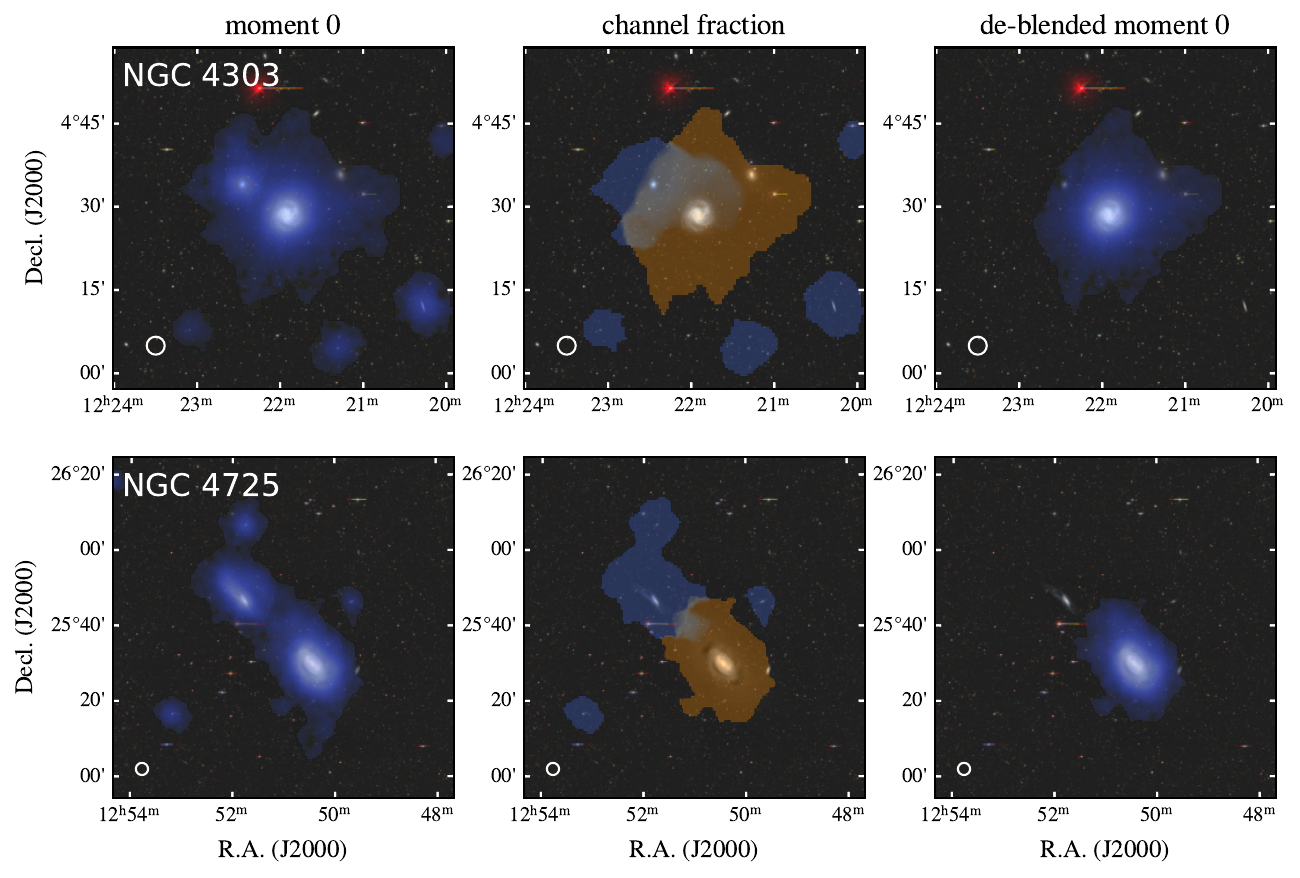}
    \caption{Examples of 3D de-blending in the merging systems. Left: moment 0 images before de-blending. Middle: the fraction of channels assigned to the target galaxy at each spaxel, from 0\% (blue), 50\% (gray) to 100\% (orange). Right: moment 0 images of the target galaxies after de-blending.
    The white circles at the bottom-left corner of each panel represent the beam size of the FEASTS observation. The optical images in the background are from the Legacy Survey \citep{Dey19}.}
    \label{fig:deblend_atlas}
\end{figure*}

In Figure \ref{fig:deblend_atlas}, we show two examples of early-stage mergers observed by FEASTS. After de-blending, the $\hi$ flux of the target galaxies is clearly separated from its neighbors in the overlapping region. 

\section{The Reliability of HI Size Measurements from FEASTS images}
\label{appendix:feasts}
We have developed procedures in previous studies to cross calibrate single-dish and interferometry data \citepalias{Wang23,Wang24}, and combine them into coherent moment-0 images \citep{Wang24b}. 
Because these combined images have the same high spatial resolution as but do not miss flux as the input interferometric images, we can accurately measure $R_1$ and $R_{001}$ from them (see justification in Appendix~\ref{appendix:combined}). 
We can downgrade the combined images and simulate FAST observations with them. 
Using the measurements from the input combined images as true values, we can test how well $R_1$ and $R_{001}$ can be measured at the resolution of FEASTS, for different true $R_1$ values.

Among the analysis sample in the main text part, there are 10 galaxies with interferometric data from the surveys HALOGAS (4) or THINGS (6), allowing us to combine them with the single-dish $\hi$ images of FEASTS. 
We take two additional single dish+interferometry combined $\hi$ images published in previous studies, NGC 2403 \citep{deBlok18}  and NGC 5236 \citep{Eibensteiner23}. 
These 12 combined images are used as the input of simulation.

Here is the procedure of simulating the FAST observation with them.
We shift the input images to different further distances, to mimic disks with smaller apparent sizes. 
Expectedly, a wide distance range is allowed for initially larger disks, if we set a minimal $R_1$ for the simulated disks.
The corresponding true $\hi$ radius are scaled accordingly and recorded.
We then convolved the shifted images with the FAST average beam \citepalias{Wang24}, to produce simulated images of FAST. 
The FAST beam has a FWHM of 3.24$'$. 

\begin{figure*} 
  \centering
  \includegraphics[width=8.cm]{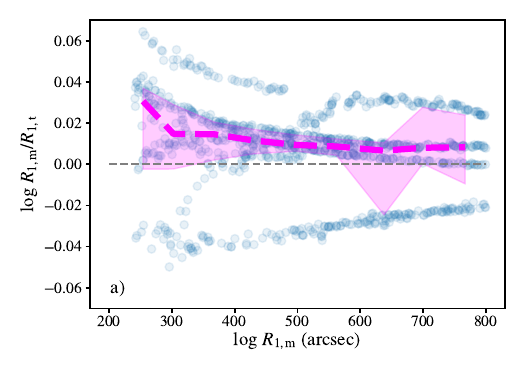}
  \includegraphics[width=8.cm]{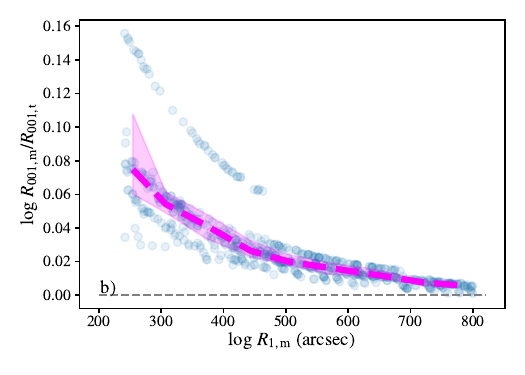}
  \includegraphics[width=8.cm]{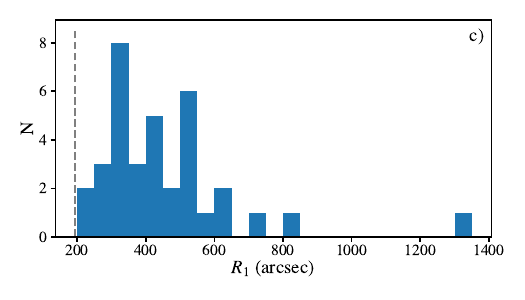}
    \caption{Simulation to test the accuracy of measuring $R_1$ and $R_{001}$ from FAST images. The top panels show the deviation from true values ($R_{x,{\rm t}}$, $x=$1 or 001) for the measurements ($R_{x,{\rm m}}$) varying as a function of $R_{\rm 1,{\rm m}}$. The data points are in light-blue dots. The median $R_{x,{\rm m}}/R_{x,{\rm t}}$ as a function of $R_{x,{\rm m}}$ is plotted in magenta dashed curves, and the 25 to 75 percentile scatters are plotted in light magenta shaded regions. The horizontal black dashed lines mark the position of zero. The bottom panel shows the $R_1$ distribution of the analyzed sample used in the main part of the paper. The vertical dashed grey line mark the FAST beam FWHM.}
  \label{fig:simu}
  \end{figure*}

We measure $R_1$ and $R_{001}$ from the simulated images ($R_{1,{\rm m}}$ and $R_{001,{\rm m}}$), and compare with the true values ($R_{1,{\rm t}}$ and $R_{001,{\rm t}}$). 
The results are presented in the top row of Figure~\ref{fig:simu}.
Most data points concentrate into a coherent trend in each panel, but there are a few outlying branches with larger $R_{\rm m}$ versus $R_{\rm t}$ differences than the other data points.
Each outlying branch corresponds to one original input galaxy, which happen to have significant small-scale structure on the radial $\SHI$ profile close to the characteristic radius.
We will need a larger sample in the future, in order to robustly quantify how the coupling of such structures with poor spatial resolution lead to systematic uncertainties in radius measurements.
So far, galaxies with uncertainties of this type in our test have an incidence rate of $3/12$ for $R_1$, but the highest systematic uncertainty does not exceed 0.065 dex.
The largest systematic uncertainty of this type is higher for $R_{001}$ (0.16 dex), but the incidence rate of corresponding galaxies is only $1/12$.    
Considering that either the uncertainties or the incidence rates are low for this type of outliers, we focus on the behavior of major trend in the following.

For the best-resolved images in this simulation, the scatter and median values of $R_{\rm m}/R_{\rm t}$ are small ($<$0.02 dex) for both $R_1$ and $R_{001}$ measurements.
The median values and scatter of $R_{\rm m}/R_{\rm t}$ for both radius increase when disks are smaller. 
When the measured $R_1=250''$ (1.28 times the FAST FWHM), the median value of $R_{1,{\rm m}}/R_{1,{\rm t}}$ and $R_{001,{\rm m}}/R_{001,{\rm t}}$ increase to 0.03 and 0.08 dex, while the corresponding scatters both increase to $\sim$0.02 dex. 

All except for two galaxies in the analysis sample have $R_1>250''$ (bottom panel of Figure~\ref{fig:simu}).
None of the conclusions change if we exclude those two smallest disks from the analysis.
On the other hand, the derived size-mass relations in Section~\ref{sec:size} have scatters comparable to the offsets (and scatter) found here. 
Consistently, in section~\ref{sec:size}, the mcmc analysis indicates the intrinsic scatter of the $R_{001}$-$\MHI$ relation to be lower than 0.001.
We thus conclude that the FEASTS images allowed reasonably accurate measure of $R_1$ and $R_{001}$, but the scatter in each size-mass relation should be significantly contributed by these measurement uncertainties.  

\subsection{Correcting for Biases and Estimating Uncertainties for $R_1$ and $R_{001}$ Measurements}

The median curves in the top panels of Figure~\ref{fig:simu} are used to correct for biases due to beam smoothing. 
Based on the directly measured $R_1$, we interpolate the median curves to obtain the smoothing-related biases of $R_1$ and $R_{001}$ with respect to true values for each galaxy. 
We estimate the error of this correction as 0.74 times the difference between the 25 and 75 percentiles (equivalent to 1-$\sigma$) at a given directly measured $R_1$. 

The uncertainties of $R_1$ and $R_{001}$ are estimated as the square root of the quadratic sum of systematic uncertainties due to smoothing correction (described above) and uncertainties due to image errors. 

The image errors include pointing errors ($\sim10''$, \citealt{Jiang19}) and noise of pixel values (propagated from the cube noise). 
We estimate the image-error-related uncertainties by simulating 100 ``perturbed'' images for each observed moment-0 image.  
We randomly shift the image with extents following a gaussian distribution with $\sigma=10''$.
We also add random noise to the pixel values following a gaussian distribution with $\sigma$ equivalent to the 1-$\sigma$ depth of image. 
We run the same $\SHI$ and size measuring procedure on these perturbed images as for the real images. 
We calculate the standard deviation of $R_1$ ($R_{001}$) distribution from the 100 perturbed images, and view it as the image-error-related size uncertainty. 

\section{The Reliability of HI size Measurements from Combined Images}
\label{appendix:combined}
To optimize the procedures that combine single-dish and interferometry images, we have developed procedures to generate mock galaxies, and simulate single-dish and interferometric observations \citepalias{Wang24}. 
We tested using different types of interferometric data reduction products as input in the combination.
The data types include the clean model convolved with the clean beam ($I_{\rm conv}$), $I_{\rm conv}$ plus rescaled clean residual ($I_{\rm rescaled}$), and $I_{\rm conv}$ plus full residual ($I_{\rm standard}$).
Please refer to \citepalias{Wang24} for a detailed discussion on pron and cons of these different types as interferometric data products alone.
They are combined with single-dish images with the standard Fourier-space linear combination method, but using improved procedure of \citet{Wang24b}, which produced the combined images $C_{\rm conv}$, $C_{\rm rescaled}$, and $C_{\rm standard}$, respectively.
By comparing measurements to true values, we were able to find the optimized way of cross-calibrating the single-dish and interferometric data is to use $I_{\rm rescaled}$ \citepalias{Wang24}, while the combined image that most accurately recover $\SHI$ near $R_1$ is $C_{\rm conv}$ \citep{Wang24b}. 

\begin{figure*} 
  \centering
  \includegraphics[width=18cm]{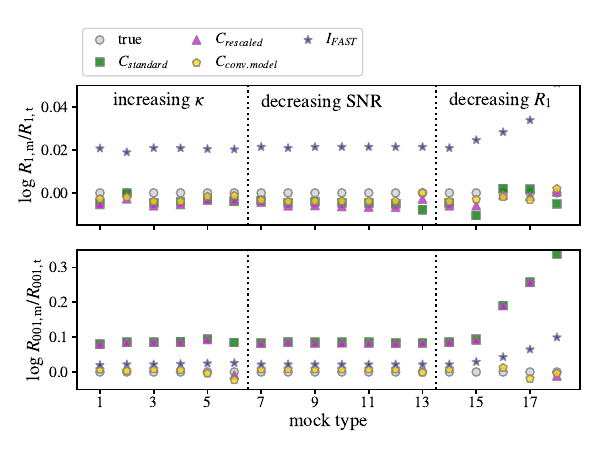}
    \caption{Mock test to select the type of combined image that best recovers $R_1$ and $R_{001}$. The different colors represent measurements from different types of images, including the combined ones $C_{\rm conv}$, $C_{\rm rescaled}$, and $C_{\rm standard}$, the FAST images ($I_{\rm FAST}$), and the true value. The different mock types are labeled by numbers, with 1--6 with increasing power spectral slopes, 7--13 with decreasing SNRs, 14-18 with decreasing $R_{1,\rm{t}}$ from 9 to 5$'$ with a step of 1$'$.  }
  \label{fig:mock}
  \end{figure*}

Here we add upon the previous tests presented in \citet{Wang24b}, and ask which type of combined image best recovers $R_1$ and $R_{001}$. 
We generate 18 sets of mock data.
The first 13 sets have a fixed $\hi$ disk size $R_1=9'$, which are the same mocks used in \citet{Wang24b}.
Sets 1 to 6 have a fixed integral signal-to-noise ratios (SNR) of 2.8, but integral increasingly larger $\hi$ disk power spectral slopes from 1.8 to 3.2.
Sets 7 to 13 have a fixed power spectral slope of 2.2, but decreasing SNR from 4.4 to 0.9. 
Sets 14 to 18 have a fixed power spectral slope of 2.2, and a fixed integral SNR of 2.8, but different $\hi$ disk sizes, with $R_{1}$ decreasing from 9$'$ to 5$'$ with a step of 1$'$.
The result is present in Figure~\ref{fig:mock}.
It is clear that from $C_{\rm conv}$, we can obtain most accurate $R_1$ and $R_{001}$ measurements, with offsets from true values less than 0.01 dex in most cases. 
There is no strong dependence on power spectral slopes of $\hi$ disks or SNRs of data for these results.

We also check and find that the measurements from FAST images are also reasonably good for these mock disks, with offset less than 0.1 dex when the $R_{1,\rm{t}}< 5'$.
Because the statistics is small, we conduct a more careful investigation on the possibility of accurately measuring sizes from FAST images in Appendix~\ref{appendix:feasts}.

The conclusion here is that we can accurately measure $R_1$ and $R_{001}$ from the combined images when the true $R_1>5'$. 

\section{Tables}
\label{appendix:size_table}
We present for the whole sample the measurements of $R_1$ and $R_{001}$, and their errors in Table~\ref{tab:size}.
We present the median profile of $\SHI$ and the errors (the data points of the black curve in Figure~\ref{fig:approx_profile}) in Table~\ref{tab:medprof}.

\begin{table}
  \centering
          \caption{$\hi$ Size Measurements}
          \begin{tabular}{c c c c c c}
            \hline
            Galaxy  & $\log \, (\frac{M_{\rm HI}}{M_{\odot}})$  &  $\log\,(\frac{R_{\rm 001}} {{\rm kpc}})$ & $\sigma_{R001}$  & $\log\,(\frac{R_{\rm 1}} {{\rm kpc}})$ & $\sigma_{R1}$   \\
                  & (dex)  & (dex) & (dex)  & (dex) & (dex) \\
            (1)     & (2)     & (3)  & (4)  & (5) & (6)   \\
            \hline
            NGC 2541 & 9.66 & 1.59 & 0.01 & 1.30 & 0.03 \\
            NGC 2841 & 10.01 & 1.91 & $<$0.01 & 1.53 & 0.03 \\
            NGC 2903 & 9.66 & 1.63 & $<$0.01 & 1.32 & 0.03 \\
            NGC 3169 & 10.37 & 2.28 & 0.01 & 1.59 & 0.03 \\
            NGC 3198 & 10.08 & 1.78 & 0.01 & 1.51 & 0.03 \\
            NGC 3338 & 10.17 & 1.87 & 0.01 & 1.55 & 0.03 \\
            NGC 3344 & 9.75 & 1.67 & 0.01 & 1.31 & 0.03 \\
            NGC 3486 & 9.88 & 1.69 & 0.01 & 1.39 & 0.02 \\
            NGC 3521 & 10.02 & 1.78 & $<$0.01 & 1.45 & 0.03 \\
            NGC 3718 & 10.05 & 1.89 & 0.01 & 1.52 & 0.03 \\
            NGC 4214 & 8.80 & 1.12 & $<$0.01 & 0.84 & 0.03 \\
            NGC 4254 & 9.85 & 1.81 & 0.01 & 1.38 & 0.03 \\
            NGC 4414 & 9.75 & 1.73 & 0.03 & 1.35 & 0.02 \\
            NGC 4449 & 9.50 & 1.68 & $<$0.01 & 1.07 & 0.03 \\
            NGC 4490 & 9.91 & 1.92 & $<$0.01 & 1.37 & 0.03 \\
            NGC 4532 & 9.63 & 1.79 & 0.04 & 1.16 & 0.03 \\
            NGC 4559 & 9.82 & 1.58 & 0.01 & 1.32 & 0.03 \\
            NGC 4725 & 9.71 & 1.70 & 0.01 & 1.39 & 0.03 \\
            NGC 5033 & 10.22 & 1.85 & 0.01 & 1.56 & 0.03 \\
            NGC 5055 & 10.06 & 1.88 & $<$0.01 & 1.55 & 0.03 \\
            NGC 5194 & 9.74 & 1.78 & 0.01 & 1.24 & 0.03 \\
            NGC 5457 & 10.37 & 1.88 & $<$0.01 & 1.65 & 0.03 \\
            NGC 628 & 10.11 & 1.78 & $<$0.01 & 1.52 & 0.03 \\
            NGC 672 & 9.49 & 1.63 & $<$0.01 & 1.38 & 0.03 \\
            NGC 864 & 9.82 & 1.65 & 0.01 & 1.39 & 0.02 \\
            NGC 4303 & 9.91 & 1.77 & 0.01 & 1.41 & 0.03 \\
            NGC 4496 & 9.45 & 1.56 & 0.04 & 1.20 & 0.01 \\
            NGC 5248 & 9.56 & 1.57 & 0.01 & 1.27 & 0.02 \\
            NGC 4395 & 9.26 & 1.31 & $<$0.01 & 1.06 & 0.03 \\
            NGC 3368 & 9.56 & 1.70 & 0.02 & 1.24 & 0.03 \\
            NGC 5474 & 9.02 & 1.25 & 0.01 & 0.99 & 0.02 \\
            SexB & 7.71 & 0.65 & 0.01 & 0.34 & 0.02 \\
            IC1727 & 9.29 & 1.73 & 0.01 & 1.38 & 0.03 \\
            NGC 4618 & 8.86 & 1.33 & 0.03 & 1.00 & 0.02 \\
            NGC 4536 & 9.70 & 1.60 & 0.02 & 1.30 & 0.02 \\
            \hline
          \end{tabular} 

            \raggedright
              Column~(1): Galaxy name.  
              Column~(2): $\hi$ mass from FEASTS data.
              Column~(3): Characteristic radius $R_{001}$ for the $\hi$ disk, measured at the $\SHI=0.01\, \Msunpcsq$ iso-density level. 
              Column~(4): Error of $R_{001}$. 
              Column~(5): Characteristic radius $R_1$ for the $\hi$ disk, measured at the $\SHI=1\, \Msunpcsq$ iso-density level. 
              Column~(6): Error of $R_1$. 
      \label{tab:size}
  \end{table}

  \begin{table}
    \centering
    \caption{$\hi$ Median profile of $\SHI$}
    \begin{tabular}{ccc}
    $r/R_{001}$ & $\log \SHI/(\Msunpcsq)$ & $\sigma(\log \SHI)$ \\
    (1) & (2) & (3) \\
    \hline
    0.00 & 0.77 & 0.03 \\
    0.05 & 0.77 & 0.03 \\
    0.10 & 0.74 & 0.03 \\
    0.15 & 0.71 & 0.03 \\
    0.20 & 0.65 & 0.03 \\
    0.25 & 0.58 & 0.04 \\
    0.30 & 0.46 & 0.04 \\
    0.35 & 0.35 & 0.04 \\
    0.40 & 0.19 & 0.05 \\
    0.45 & 0.03 & 0.05 \\
    0.50 & -0.15 & 0.06 \\
    0.55 & -0.33 & 0.06 \\
    0.60 & -0.51 & 0.06 \\
    0.65 & -0.68 & 0.06 \\
    0.70 & -0.87 & 0.06 \\
    0.75 & -1.05 & 0.05 \\
    0.80 & -1.23 & 0.05 \\
    0.85 & -1.44 & 0.06 \\
    0.90 & -1.63 & 0.08 \\
    0.95 & -1.81 & 0.11 \\
    1.00 & -2.00 & 0.17 \\
    1.05 & -2.19 & 0.27 \\
    1.10 & -2.39 & 0.42 \\
    1.15 & -2.58 & 0.66 \\
    1.20 & -2.79 & 1.06 \\
    \hline
  \end{tabular}

    \raggedright
    Column~(1): Radius in unit of $R_{001}$.  
    Column~(2): $\SHI$ values of the $\SHI$ median profile.
    Column~(3): Errors of the $\SHI$ median profile.
  \label{tab:medprof}
  \end{table}

\bibliography{fullprof}{}
\bibliographystyle{apj}

\end{document}